# Mathematical model of solid tumor formation

R.G.Khlebopros[1], V.A.Slepkov[2], V.G.Sukhovolsky[2], Y.V.Mironov[2,3], V.E.Fedorov[3], S.P.Gabuda[3*]

[1]*Institute of Biophysics, KSC SB RAS, Akademgorodok, Krasnoyarsk, 660036, Russia*
[2] *International Centre for Critical States Research, KSC SB RAS, Akademgorodok, Krasnoyarsk, 660036, Russia*
[3]*Nikolayev Institute of Inorganic Chemistry of Siberian Department of Russian Academy of Sciences. Novosobirsk 630090, Russian Federation*

\_\_\_\_\_\_\_\_\_\_\_\_\_\_\_\_\_\_

*) Corresponding author: gabuda@casper.che.nsk.su

**Abstract.** The problem of the onset and growth of solid tumour in homogeneous tissue is regarded using an approach based on local interaction between the tumoral and the sane tissue cells. The characteristic sizes and growth rates of spherical tumours, the points of the beginning and the end of spherical growth, and the further development of complex structures (elongated outgrowths, dendritic structures, and metastases) are derived from the assumption that the reproduction rate of a population of cancer cells is a non-monotone function of their local concentration. The predicted statistical distribution of the characteristic tumour sizes, when compared to the clinical data, will make a basis for checking the validity of the theory.

Key words: tumour growth, mathematical model, local interaction

## 1. Introduction

Tumour growth modelling has been the subject of much recent literature (e.g., [1-12]), but the existing mathematical models appear to neglect the local interaction between cancer and tissue cells. This interaction between cancer and tissue cells affects the cancer cells' reproduction rate the way that it becomes dependent on their local concentration [13-17]. Therefore, tumour growth is mainly determined by the behaviour of cancer cells adjacent to the tumour surface. This process, in its turn, is largely dependent on the tumour's local curvature. We apply this approach to investigate the macroscopic effects of tumour growth in a homogeneous tissue. The effects of local interaction can be most simply described using the reproduction function approach. Figure 1 shows hypothetical non-monotone curves 1-3 describing the reproduction rate of cancer cells $\dot{v}$ vs. local concentration $v = \frac{x}{x+y}$, where $x$ and $y$ are the number of cancer and tissue cells, respectively, in the considered tissue area. (For simplicity and without loss of generality, cancer and tissue cells are assumed hereafter to be of the same size). The curves represent the interaction of carcinogenic factors with the immune system, the availability of nutrients, and the local cellular interaction. On each plot, the first interval of decrease corresponds to the activation of the immune system in response to the presence of cancer cells in the organism. The same behaviour is expected when the immune system in affected organism is supported by definite antitumoral medicines. Among them are widely known antioxidants like selenium. The series of methods as aimed in direct damage of cancer cells. The newest is the use of some radioactive-isotope containing complexes employed for treatment of different cancer diseases - from kidney, lung and stomach to brain and bones [18-21]. Very recently, in medical practice is widely used the photo dynamic therapy (PDT) based on the luminescence of introduced preparation which can activate oxygen and destroy the tumor cells. We suppose that this phenomenon is of special interest for rhenium complexes: the combination of metal cluster luminescence and of metal radioactivity (the $^{186}$Re and $^{188}$Re radioactive isotopes) can create synergetic effect in cancer therapeutics. The use of similar improved methods for tumour treatment need in detailed modelling of the tumour growth.

Furthermore, as the concentration of cancer cells increases, the suppression of their reproduction through local interaction between cancer and tissue cells is first slowed, then completely stopped. Finally, it gives way to a reverse process in which cancer cells stimulate the pathological division of the tissue cells [14, 15]. On the plots, this process corresponds to the interval of increase. A still higher concentration of cancer cells causes a deficit of nutrients which results in the inhibition of cancer cells' reproduction rate and finally to their death. On the plots, this process corresponds to the second interval of decrease.

Curve 1 describes the case when tumour onset in the tissue is fully suppressed. When the immune system cannot provide complete suppression (curve 2), tumours can spontaneously arise and develop. This case is the subject of our study. Curve 3 images a suppressed immune system when the tumour spreads throughout the organism.

## 2. Growth and formation of solid tumours

For the sake of simplicity we are solving this problem in two dimensions which corresponds to the behaviour of tumour sections. The difference between the three- and two-dimensional formulations is of minor importance as we consider the processes only qualitatively. Assume that the reproduction function is given by $\dot{v} = \varphi(v)$, $R$ is the radius of a spherical tumour, and $r$ is the radius of cancer and tissue cells. Define the density $v$ of cancer cells in the vicinity of the surface as $\frac{S_c}{S_c + S_t}$, where $S_c = 4\pi r(R-r)$ is the area of a layer of tumour cells that contact the surface from inside, and $S_t = 4\pi r(R+r)$ is the area of a layer of tissue cells that contact the surface from outside:

$$v = \frac{R-r}{2R}. \tag{1}$$

$R$ and $\dot{R}$ are found from (1) as

$$R = \frac{r}{1-2v}, \tag{2}$$

$$\dot{R} = \frac{2r\dot{v}}{(1-2v)^2}. \tag{3}$$

Let a tumour occupy the area $V = 2\pi R$. Then

$$\dot{V} = 2\pi R \dot{R}. \tag{4}$$

On the other hand, $\dot{V}$ is the area of cancer cells produced per unit time in the area $S_c + S_t$. Therefore

$$\dot{V} = (S_c + S_t)\varphi(v) = 8\pi r R \varphi(v), \tag{5}$$

From (4) and (5) obtain

$$\dot{R} = 4r\varphi(v).$$

Substituting the latter equation in (3) obtain differential equation for the density of cancer cells in the tumour's surface layer:

$$\dot{v} = 2\varphi(v)(1-2v)^2 = \psi(v). \tag{6}$$

Figure 2 shows the plot of $\psi(v)$ (solid line) compared with the plot of $\varphi(v)$ (dashed line).

From (2) find the minimum ($R_{\min}$) and the maximum ($R_{\max}$) radiuses of a spherical tumour and the critical value ($R_t$) above which the tumour begins to grow until it reaches $R_{\max}$:

$$R_{\min} = R_1 = \frac{r}{1-2v_1}, \quad R_t = R_2 = \frac{r}{1-2v_2}, \quad R_{\max} = R_3 = \frac{r}{1-2v_3}.$$

Radius $R_0 = R(v_0)$ ($v_0$ is the maximum point) corresponds to the highest growth rate of cancer cells in the tumour's surface layer.

Between $R_2$ and $R_0$ the spherical shape of the tumour is stable against minor deformations of the surface curvature. When the sphere is extended into an ellipsoid, the areas with a smaller curvature radius tend to grow more slowly while the areas with a larger curvature radius grow faster, and the tumour thus recovers its spherical shape (Fig. 3a). When the initial tumour radius is between $R_0$ and $R_3$, the spherical shape of the tumour becomes unstable as the growth rate is inversely proportional to the curvature radius – and the tumour becomes extended into an ellipsoid (Fig. 3b). The further evolution can follow several scenarios. As long as the curvature radius at the sharp end of the ellipsoid exceeds $R_0$ it remains the site of the highest reproduction rate and extends more and more transforming the tumour into an elongated outgrowth (Fig. 3c). When the curvature radius at the sharp end of the outgrowth becomes smaller than $R_0$, the sites of fastest reproduction remove along the sides of the stem to the points of curvature radius $R = R_0$ which causes new outgrowths and transforms the tumour into a dendritic structure (Fig. 3d). Later on the new outgrowths may detach from the main stem in the region of negative curvature corresponding to a negative growth of cancer cells (Fig. 3e). If a negative curvature develops at the junction of the outgrowth with the main tumour, the outgrowth may also detach and evolve as a separate metastasis (Fig. 3f).

## 3. Statistical distribution of tumours
### 3.1. General case
First consider the statistical distribution of tumours in a general case. Let the spontaneous onset of a tumour with the radius $R = R(v)$ in the tissue be given by the probability density $\eta(v)$, and the rate of density growth in the tumour's surface layer be given by

$$\dot{v} = \psi(v), \tag{7}$$

where $v$ is between $v_1$ and $v_2$. Find the average probability density $p(v)$ which is the distribution of tumours in the tissue in the time $T$ elapsed after the beginning of the process. Let the general solution of (7) be

$$v = \bar{v}(t, v_0), \tag{8}$$

where $v_0$ is $v$ at the initial time $t = 0$ and $v_1 \leq v_0 \leq v_2$. From (8) obtain

$$dv = \frac{\partial \bar{v}}{\partial v_0} dv_0 \tag{9}$$

for each fixed moment $t$. As the interval $dv$ is the image of the interval $dv_0$, the probabilities of these intervals are equal:

$$\eta(v) dv = \eta(v_0) dv_0, \tag{10}$$

where $v_1 \leq v \leq v_2$, $v_1 \leq v_0 \leq v_2$ and $v = v(t, v_0)$. From (9) and (10) obtain

$$\eta(v) \frac{\partial \bar{v}}{\partial v_0} dv_0 = \eta(v_0) dv_0. \tag{11}$$

Assume that at each moment $t$ the function $v$ has the reverse function

$$v_0 = \bar{v}_0(v, t),$$

and rewrite (11) as

$$\eta(v) = \frac{\partial \bar{v}_0}{\partial v} \eta(v_0). \tag{12}$$

Then the equation for the probability density at the point $v$ at the moment $t$ is

$$p(v,t) = \begin{cases} \dfrac{\partial \bar{v}_0}{\partial v} \eta(\bar{v}_0(v,t)), & v \in [\bar{v}(t,v_1), \bar{v}(t,v_2)] \\ 0, & v \notin [\bar{v}(t,v_1), \bar{v}(t,v_2)] \end{cases}. \tag{13}$$

The probability $p(v)$ is found by integrating $p(\hat{i}, t)$ with respect to $t$ over 0 to $T$:

$$p(v) = \frac{1}{T} \int_0^T p(v,t) dt, \quad v_1 \leq v \leq v_2. \tag{14}$$

*3.2. Specific model*

As only a qualitative description is supposed, specify $\eta(v)$ and $\psi(v)$ as the simplest functions with the required properties. Thus the function $\eta(v)$ (the probability density that a tumour with the radius $R = \dfrac{r}{1-2v}$ arises spontaneously in the tissue) must reach its maximum at zero and decrease rapidly. Assume that

$$\eta(v) = e^{-v}. \tag{15}$$

Note that the density $v = 0$ corresponds to the radius $R = r$ and $v = \tfrac{1}{2}$ corresponds to $R = \infty$. Let $\psi(v)$ be represented by a piecewise linear function $\dot{v} = \chi(v)$ (Fig. 4.), assuming for simplicity and without loss of generality that

$$v^1 = \frac{v_1 + v_2}{2}, \quad v^2 = \frac{v_2 + v_3}{2}, \quad v^3 = \frac{v_3 + \tfrac{1}{2}}{2}.$$

Then denote

$$\Delta v^1 = v_2 - v_1, \quad \Delta v^2 = v_3 - v_2, \quad \Delta v^3 = 1/2 - v_3,$$

$$\alpha = \frac{\alpha^0}{v_1} = \frac{2\alpha^1}{\Delta v^1} = \frac{2\alpha^2}{\Delta v^2} = \frac{2\alpha^3}{\Delta v^3}.$$

Besides, denote $\varepsilon = e^{-\alpha T}$. As we consider only large intervals $T$, $\varepsilon$ is implied to be small enough.
Find $p(v)$ separately at the intervals $[0, v_1]$, $[v_1, v_2]$, and $[v_2, v_3]$.

1) Let $0 \leq v \leq v_1$.

Here $\dot{v} = -\alpha(v - v_1)$, and the general solution is

$$v = v_1 + (v_0 - v_1)e^{-\alpha t}, \qquad 0 \leq v_0 \leq v_1. \tag{16}$$

From (16) express $v_0$ through $v$:

$$v_0 = e^{\alpha t}v - v_1(e^{\alpha t} - 1), \quad v_1(1 - e^{-\alpha t}) \leq v \leq v_1.$$

Note that $\dfrac{\partial v_0}{\partial v} = e^{\alpha t}$. Then (13) gives

$$p(v,t) = \begin{cases} e^{-v_1} e^{\alpha t} e^{(v_1 - v)e^{\alpha t}}, & v_1(1 - e^{-\alpha t}) \leq v \leq v_1 \\ 0, & 0 \leq v \leq v_1(1 - e^{-\alpha t}) \end{cases}.$$

From (14) obtain:

$$p(v) = \begin{cases} \dfrac{e^{-v_1}}{T}\displaystyle\int_0^{t(v)} e^{\alpha t} e^{(v_1-v)e^{\alpha t}}\,dt, & 0 \leq v \leq v_{1T} \\[2mm] \dfrac{e^{-v_1}}{T}\displaystyle\int_0^{T} e^{\alpha t} e^{(v_1-v)e^{\alpha t}}\,dt, & v_{1T} \leq v \leq v_1 \end{cases}.$$

Where $t(v) = -\dfrac{1}{\alpha}\ln\left(\dfrac{-v + v_1}{v_1}\right)$ is the reverse function of (16). Since $\int e^{\alpha t} e^{(v_1-v)e^{\alpha t}}\,dt = \dfrac{e^{(v_1-v)e^{\alpha t}}}{\alpha(v_1 - v)}$, finally obtain:

$$p(v) = \begin{cases} \dfrac{e^{-v_1}}{\alpha T}\dfrac{e^{v_1} - e^{v_1-v}}{v_1 - v}, & 0 \leq v \leq v_{1T} \\[2mm] \dfrac{e^{-v_1}}{\alpha T}\dfrac{e^{(v_1-v)e^{\alpha T}} - e^{v_1-v}}{v_1 - v}, & v_{1T} \leq v \leq v_1 \end{cases}.$$

2) Let $v_1 \leq v \leq v_2$.
Then

$$\dot{v}(v) = \begin{cases} \alpha(-v+v_1), & v_1 \leq v \leq v^1 \\ \alpha(v-v_2), & v^1 \leq v \leq v_2 \end{cases}. \tag{17}$$

Write general solutions separately for the two intervals:

$$v(v_0,t) = v_0 e^{-\alpha t} + v_1(1-e^{-\alpha t}), \quad v_1 \leq v \leq v^1 \tag{18'}$$

$$v(v_0,t) = v_0 e^{\alpha t} + v_2(1-e^{\alpha t}), \quad v^1 \leq v \leq v_2 \tag{18''}$$

Then write the general solution of (17):

$$v(v_0,t) = \begin{cases} v_0 e^{-\alpha t} + v_1(1-e^{-\alpha t}), & v_1 \leq v_0 \leq v^1 \\ v^1 e^{-\alpha(t-t_0)} + v_1(1-e^{-\alpha(t-t_0)}), & v^1 \leq v_0 \leq v_2, \ v_1 \leq v \leq v^1, \\ v_0 e^{\alpha t} + v_2(1-e^{\alpha t}), & v^1 \leq v \leq v_2 \end{cases} \tag{19}$$

where $t_0$ is the time of transition from $v_0$ to $v^1$ obtained from (18'') by substituting $v^1$ for $v$:

$$v^1 = -e^{\alpha t_0}(v_2 - v_0) + v_2,$$

$$e^{\alpha t_0} = \frac{v_2 - v^1}{v_2 - v_0} = \frac{\Delta v^1}{2(v_2 - v_0)}.$$

Then (19) looks as:

$$v(v_0,t) = \begin{cases} v_0 e^{-\alpha t} + v_1(1-e^{-\alpha t}), & v_1 \leq v_0 \leq v^1 \\ \dfrac{(\Delta v^1)^2 e^{-\alpha t}}{4(v_2 - v_0)} + v_1, & v^1 \leq v_0 \leq v_2, \ v_1 \leq v \leq v^1. \\ v_0 e^{\alpha t} + v_2(1-e^{\alpha t}), & v^1 \leq v \leq v_2 \end{cases}$$

Express $v_0$ via $v$:

$$v_0(v,t) = \begin{cases} ve^{\alpha t} - v_1(e^{\alpha t}-1), & v_1 \leq v_0 \leq v^1 & (20') \\ v_2 - \dfrac{(\Delta v^1)^2 e^{-\alpha t}}{4(v-v_1)}, & v^1 \leq v_0 \leq v_2, \ v_1 \leq v \leq v^1 & (20'') \\ ve^{-\alpha t} + v_2(1-e^{-\alpha t}), & v^1 \leq v \leq v_2 & (20''') \end{cases}$$

The constraint $v_1 \leq v_0 \leq v^1$ transforms into $v_1 \leq ve^{\alpha t} - v_1(e^{\alpha t}-1) \leq v^1$ or $v_1 \leq v \leq v_1 + \dfrac{\Delta v^1}{2}e^{-\alpha t}$. The constraints from (20'') transform into $v_1 + \dfrac{\Delta v^1}{2}e^{-\alpha t} \leq v \leq v^1$.

$$v_0(v,t) = \begin{cases} ve^{\alpha t} - v_1(e^{\alpha t}-1), & v_1 \leq v \leq v_{2T} & (a) \\ v_2 - \dfrac{(\Delta v^1)^2 e^{-\alpha t}}{4(v-v_1)}, & v_{2T} \leq v \leq v^1 & (b) \\ ve^{-\alpha t} + v_2(1-e^{-\alpha t}), & v^1 \leq v \leq v_2 & (c) \end{cases} \tag{21}$$

where $v_{2T} = v_1 + \dfrac{\Delta v^1}{2}\varepsilon$.

Integration over the areas (a), (b), (c) of the equations for $p(v,t)$ obtained from (12), (13), and (15) gives the following $p(v)$:

$$p(v) = \begin{cases} \dfrac{e^{-v_1}}{\alpha T} \dfrac{e^{\frac{v_1-v}{\varepsilon}} - e^{v_1-v}}{v_1 - v}, & v_1 \leq v \leq v_{2T} \\[2ex] \dfrac{e^{-v_1}}{\alpha T} \dfrac{e^{-(v-v_1)} - e^{-\Delta v^1} e^{\frac{\beta}{v-v_1}}}{v - v_1}, & v_{2T} \leq v \leq v^1 \\[2ex] \dfrac{e^{-v_2}}{\alpha T} \dfrac{e^{v_2-v} - e^{\varepsilon(v_2-v)}}{v_2 - v}, & v^1 \leq v \leq v_2 \end{cases}$$

In a similar manner for the interval $v_2 < v < v_3$ obtain:

$$p(v) = \begin{cases} \dfrac{e^{-v_2}}{\alpha T} \dfrac{e^{-\varepsilon(v-v_2)} - e^{-(v-v_2)}}{v - v_2}, & v_2 \leq v \leq v^2 \\[2ex] \dfrac{e^{-v_3}}{\alpha T} \dfrac{e^{\Delta v^2} e^{-\frac{(\Delta v^2)^2}{4(v_3-v)}\varepsilon} - e^{v_3-v}}{v_3 - v}, & v^2 \leq v \leq v_{3T} \\[2ex] \dfrac{e^{-v_3}}{\alpha T} \dfrac{e^{\frac{v_3-v}{\varepsilon}} - e^{v_3-v}}{v_3 - v}, & v_{3T} \leq v \leq v_3 \end{cases}.$$

Figure 5 shows the plot of function $p(v)$ for large $T$. At the points of stable equilibrium $v_1$ and $v_3$ there are two marked peaks. Note that the actual number of spherical tumours with unstable shapes over $[v_0, v_3]$ is much lower than it would be expected from the above calculations which assumed stability over the whole range of intervals.

**4. Conclusion**

The reproduction rate of cancer cells is controlled by the local concentrations of interacting cancer and tissue cells, and hence by the curvature of the surface between the tumour and the host tissue. The local concentration of cancer cells increases with the curvature radius. Assumption of non-monotone dependence of reproduction rate vs. local concentration of cancer cells reveals critical values $R_{\min}$ (corresponding to the beginning of spherical tumour growth) and $R_{\max}$ (corresponding to the end of growth). The spherical shape is stable until the radius reaches the value $R_0$ corresponding to the maximum rate of tumour growth. When the size of the tumour is between $R_0$ and $R_{\max}$, its spherical shape becomes unstable as a minor casual deformation triggers the faster reproduction of cancer cells at the sites where the curvature radius approaches $R_0$. Then the tumour transforms into an ellipsoid, elongated outgrowths, or a branching structure, which occasionally cause the detachment of metastases at the sites of negative curvature. As tumours grow at different rates through their evolution stages, it is possible to estimate the statistical distribution of their sizes and shapes.

We wish to thank Yu. Sharovskaya who drew R. Kh's attention to the role of local cellular interactions in the reproduction of cancer cells.

**Figure captions**
Fig. 1. Reproduction rates of cancer cells $\dot{v}$ as a function of their local concentration v (curves 1-3).
Fig. 2. Derivative of cancer cells density at the tumour surface $\psi(v)$ (solid line), and initial reproduction function $\varphi(v)$ (dashed line).
Fig. 3. Growth and formation of tumour at minor deformations. Dashed line shows a deformed tumour, solid line shows tumour shape some time after deformation. Arrows show the magnitude and direction of tumour growth in the corresponding points. (a) The initial tumour has a curvature radius between $R_2$ and $R_0$. (b) The initial tumour has a curvature radius between $R_0$ and $R_3$. (c) Curvature radius on the end of the "ellipsoid" exceeds $R_0$. (d) Curvature radius at the end of the outgrowth is less than $R_0$. (e) Detachment of the outgrowth in a site of negative curvature. (f) Detachment of the outgrowth from the main tumour and development of a metastasis.
Fig. 4. Piecewise linear function $\dot{v} = \chi(v)$ used as approximation for $\psi(v)$.
Fig. 5. Probability density describing statistical distribution of tumours in a tissue.

FIGURE 1

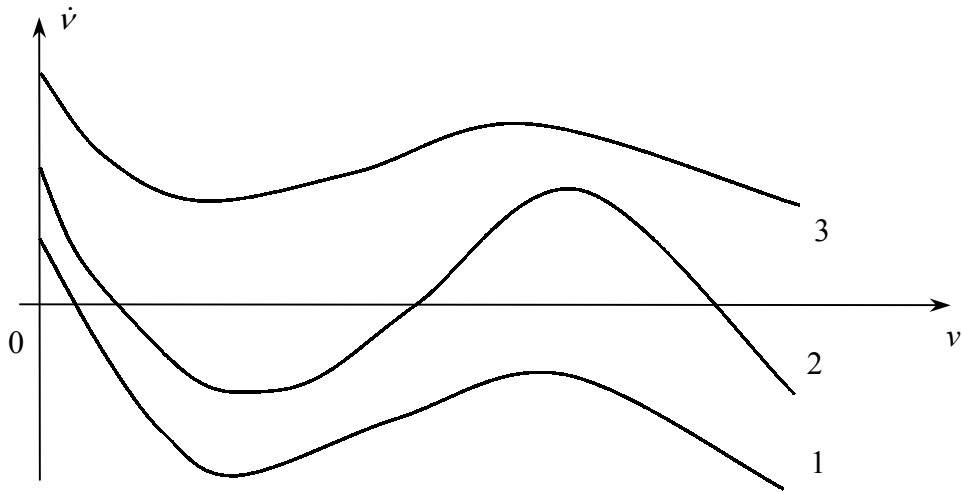

FIGURE 2

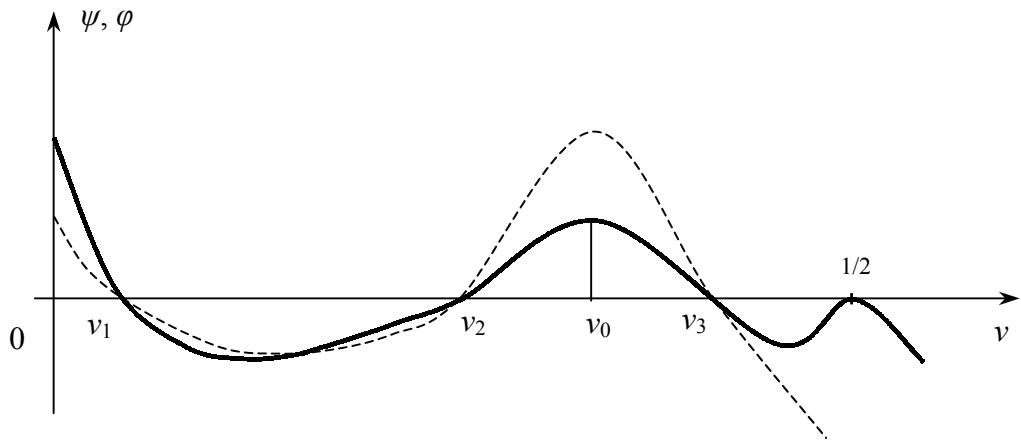

FIGURE 3a

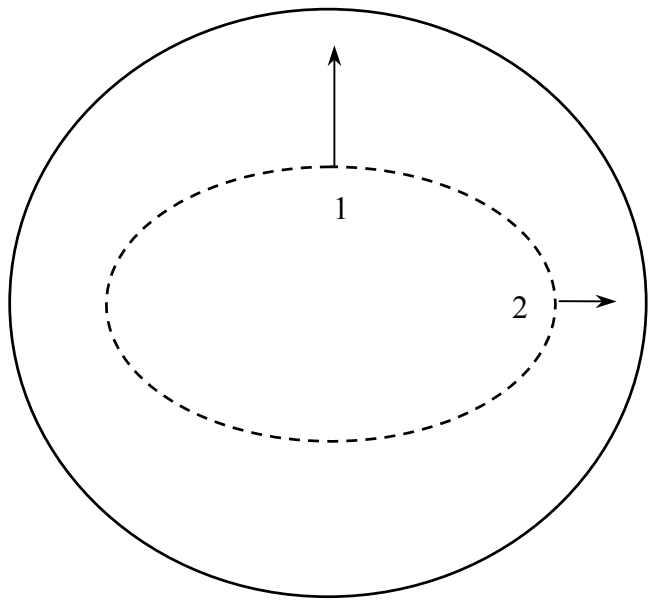

FIGURE 3b

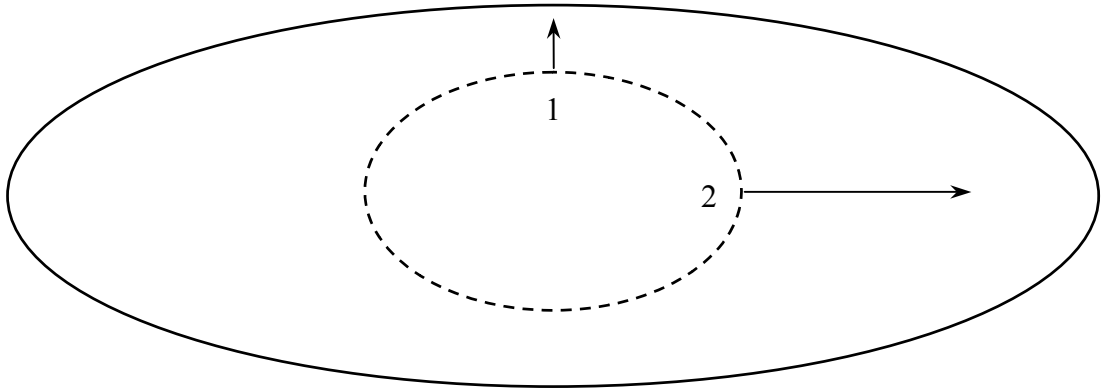

FIGURE 3c

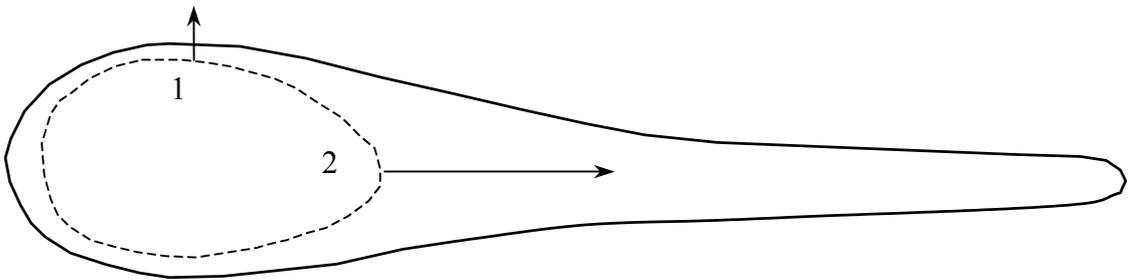

FIGURE 3d

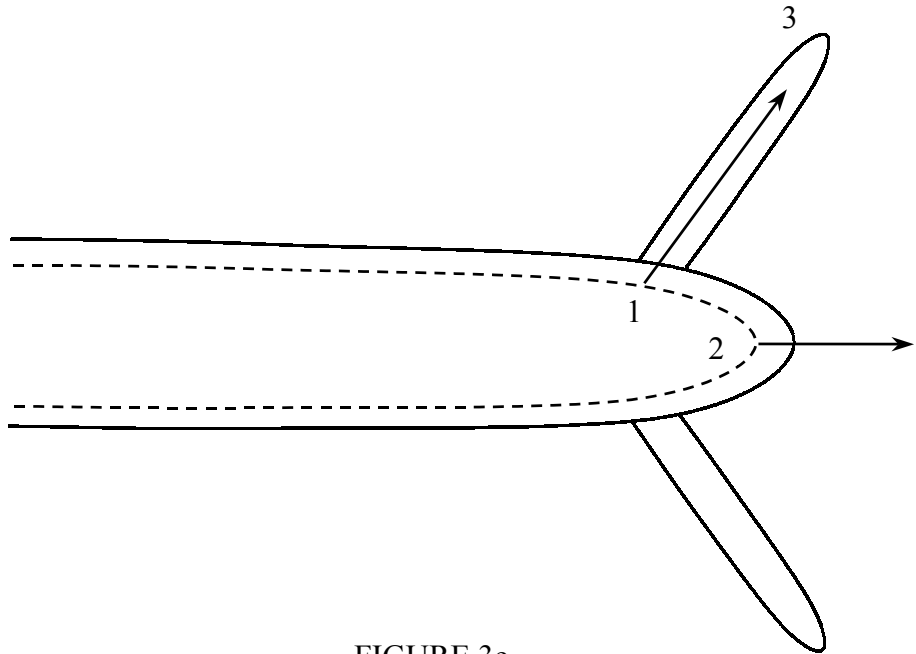

FIGURE 3e

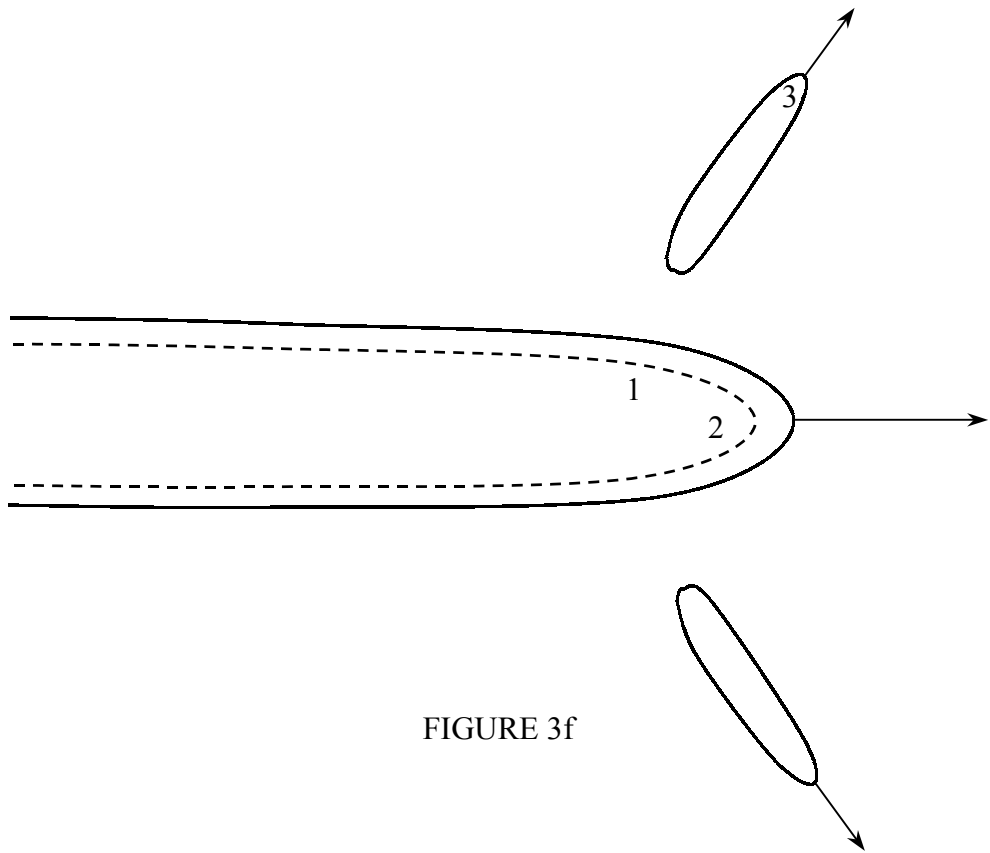

FIGURE 3f

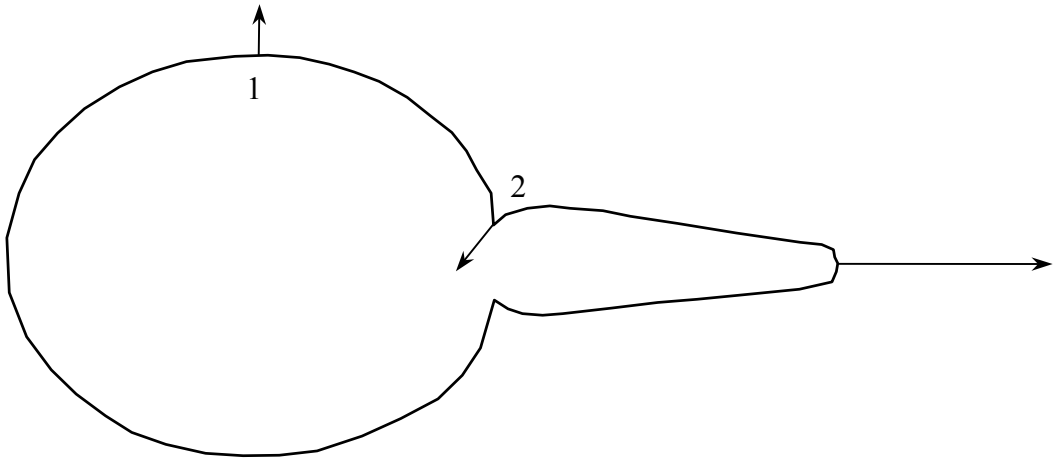

FIGURE 4

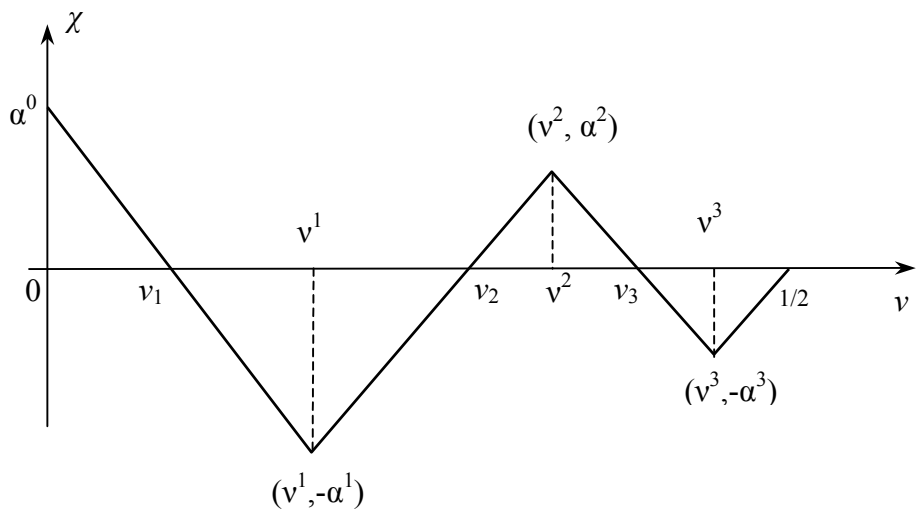

FIGURE 5

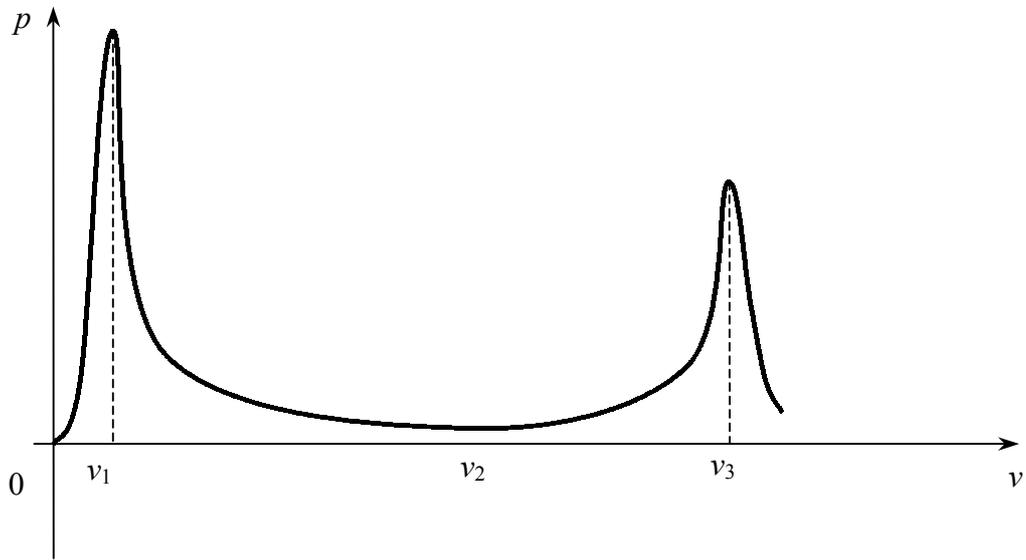